\documentclass[12pt]{article}
\DeclareMathAlphabet{\scr}{U}{rsfs}{m}{n}
\usepackage{latexsym}
\usepackage[mathscr]{eucal}
\usepackage{amsfonts}
\usepackage{amscd}
\usepackage{cite}
\usepackage{amssymb}
\usepackage[centertags]{amsmath}
\usepackage{dsfont}
\usepackage{yfonts}
\usepackage{enumerate}
\usepackage{graphicx}

\newcommand{\cleqn}{\setcounter{equation}{0}}
\newcommand{\newc}{\newcommand}
\newc{\be}{\begin{equation}}
\newc{\ee}{\end{equation}}
\newc{\bea}{\begin{eqnarray}}
\newc{\eea}{\end{eqnarray}}
\newc{\ol}{\overline}
\newc{\wt}{\widetilde}
\newc{\bs}{\boldsymbol}
\newc{\m}{\mathcal}

\begin{document}

\title{\hfill ~\\[-30mm]
       \hfill\mbox{\small SHEP-09-08}\\[30mm]
       \textbf{A~new~family~symmetry~for~{$\bs{SO(10)}$}~GUTs}}
\date{}
\author{\\Stephen F. King\footnote{E-mail: {\tt king@soton.ac.uk}}~~and
        Christoph Luhn\footnote{E-mail: {\tt christoph.luhn@soton.ac.uk}}\\ \\
  \emph{\small{}School of Physics and Astronomy, University of Southampton,}\\
  \emph{\small Southampton, SO17 1BJ, United Kingdom}}

\maketitle

\begin{abstract}
\noindent We argue that the projective special linear group
$PSL_2(7)$, also known as $\Sigma (168)$, has unique features which make it
the most suitable 
discrete family symmetry for describing quark and lepton masses
and mixing in the framework of $SO(10)$ type unified models. In
such models flavon fields in the sextet representation of
$PSL_2(7)$ play a crucial role both in obtaining tri-bimaximal
neutrino mixing as well as in generating the third family charged
fermion Yukawa couplings. In preparation for physical
applications, we derive the triplet representation of $PSL_2(7)$
in the basis $S,T,U,V$ where $S,T,U$ are the familiar triplet
generators of $S_4$ in the diagonal charged lepton basis where $T$
is diagonal. We also derive an analogous basis for the real sextet
representation and identify the vacuum alignments which lead to
tri-bimaximal neutrino mixing and large third family charged
fermion Yukawa \mbox{couplings}.
\end{abstract}
\thispagestyle{empty}
\vfill
\newpage
\setcounter{page}{1}


\section{Introduction}
\cleqn

It has been one of the long standing goals of theories of particle
physics beyond the Standard Model (SM) to predict quark and lepton
masses and mixing. With the discovery of neutrino masses and
mixing, this quest has received a massive impetus. Indeed, perhaps
the greatest advance in particle physics over the past decade has
been the measurement of neutrino masses and mixing involving two large
mixings associated with atmospheric and solar neutrino oscillation
experiments, while the remaining mixing, although unmeasured, is
constrained by reactor neutrino oscillation
experiments to be relatively small. The largeness of the two large
lepton mixings contrasts sharply with the smallness of
quark mixing, and this observation, together with the
smallness of neutrino masses, provides new and tantalising clues
in the search for the origin of quark and lepton flavour.

It turns out that the observed neutrino oscillation parameters are consistent with a
tri-bimaximal (TB) mixing structure \cite{HPS}:
\begin{equation}
U_{TB}= \left[
\begin{array}{ccc}
\sqrt{\frac{2}{3}} & \sqrt{\frac{1}{3}} & 0 \\
-\sqrt{\frac{1}{6}} & \sqrt{\frac{1}{3}} & -\sqrt{\frac{1}{2}} \\
-\sqrt{\frac{1}{6}} & \sqrt{\frac{1}{3}} & \sqrt{\frac{1}{2}}
\end{array}\right]  \ .
 \label{eq:HPS}
\end{equation}
It has been observed that this simple form might be a hint of an
underlying family symmetry $G_f$, which preserves certain group transformations
(corresponding to the generators $S,U$, as discussed later) which are
respected by  tri-bimaximal neutrino
mixing, and several models have been
constructed that account for this structure of leptonic mixing
\cite{S3-L,Dn-L,A4-L,S4-L,delta54-L}. It is possible to extend the underlying
family symmetry to provide a description of the complete fermionic structure
\cite{S3-LQ,Dn-LQ,Q6-LQ,A4-LQ,SU(5)models,A4-LQ:Morisi,A4-LQ:King,doubleA4-LQ,S4-LQ,S4-LQ:Hagedorn,Z7Z3-LQ,T7-LQ,delta27-LQ:King,SO(3)-LQ:King,SU(3)-LQ:Ross}\footnote{
See \cite{Reviews} for review papers with more extensive references on
neutrino models}, in which, in contrast to the neutrinos, the
quarks have a strongly hierarchical structure with small mixing.

A desirable feature of a complete model of quark and lepton masses
and mixing is that it should be consistent with an
underlying Grand Unified Theory (GUT) structure, either at the field theory
level or at the level of the superstring. The most ambitious
$G_f~\otimes$~GUT models which have been built to achieve this are
based on an underlying $SO(10)$ type GUT structure. This is very
constraining because it requires that all the 16 spinor components of a single family
should have the same family charge, comprising the left-handed fermions $\psi$
and the CP conjugates of the right-handed fermions $\psi^c$,
including the right-handed neutrino. We emphasise that it is very desirable
that the right-handed neutrinos should be unified with the rest
of the quarks and leptons, since this feature leads unavoidably to neutrino masses.
By contrast, in $SU(5)$ right-handed neutrinos are gauge singlets without which
it is natural to have zero neutrino masses. Although it seems somewhat {\it ad hoc}
to add right-handed neutrino singlets to $SU(5)$, once they are present
it is straightforward to construct models
of lepton masses and TB mixing that are consistent with quark masses and mixing,
and there are several successful models of this kind~\cite{SU(5)models}.
Nevertheless, it is worth remembering that in such models the right-handed neutrinos
must be added by hand as an optional extra and therefore
neutrino mass is not strictly a prediction
of $SU(5)$ whereas in $SO(10)$ neutrino mass is unavoidable
since the right-handed neutrinos are unified with the rest of the quarks and leptons,
leading to the prediction of three right-handed neutrinos.
In fact similar comments apply to any partially unified model in which
$SU(2)_R$ is gauged, for example Pati-Salam; also a gauged $U(1)_{B-L}$
  predicts right-handed neutrinos due to anomaly considerations.

Despite the theoretical attractiveness of a $G_{f}~\otimes$ GUT
structure of the form $G_{f}\otimes SO(10)$, there are very few models in this class that are capable of predicting, or more
accurately postdicting, TB mixing. As we shall discuss later, in order to obtain TB mixing naturally
(i.e. enforced by a symmetry) one requires
a group with a non-Abelian structure containing triplet representations,
together with a particular set of discrete group elements which embody the
symmetry properties of the TB mixing matrix in Eq.~(\ref{eq:HPS}).
There is also an additional theoretical requirement in $SO(10)$ that the family symmetry
must contain complex triplet representations as discussed below.
Examples of such models based on the Pati-Salam subgroup of $SO(10)$
have been constructed where the family group $G_{f}$ is $SU(3)_{f}$ \cite{SU(3)-LQ:Ross} or
$\Delta(27)$ \cite{delta27-LQ:King}.
In such models all 16 components of a single family including the left-handed fermions
$\psi$ and the CP conjugates of the right-handed fermions $\psi^c$ are assigned to the complex triplet
representation of $SU(3)_f$ or $\Delta(27)$, namely $\psi_i \sim {\bf 3},
\psi^c_i \sim {\bf 3}$. Flavon anti-triplets $\bar{\phi}^i \sim {\bf
  \overline{3}}$ are introduced to break  $G_f$, and the lowest dimension
Yukawa operators  consistent with an $SU(3)_f$ family symmetry are then,
\be
\frac{y}{M^{2}} \bar{\phi}^{i}\psi^{}_{i}\bar{\phi}^{j}\psi_{j}^{c}H \ ,
\label{Yuk_op}
\ee
where $H$ is the Higgs field which is generally taken to be a singlet $H\sim
{\bf 1}$ under $G_f$. For discrete family symmetries like $\Delta(27)$
additional contractions may in principle become possible, however, in concrete
models they are usually assumed to be suppressed or absent.
If a particular flavon $\phi_3$ has a vacuum expectation value (VEV) aligned
along the third direction $\langle \phi_3 \rangle = (0,0,V)$, then this
operator would imply a Yukawa matrix which is suitable for describing the
large third family Yukawa coupling,
\begin{equation}
Y = \left(
\begin{array}{ccc}
0 & 0 & 0 \\
0 & 0 & 0 \\
0 & 0 & \frac{yV^2}{M^2}
\end{array}
\right).  \label{Yuk}
\end{equation}

On the other hand, if the triplet representations were taken to be real rather
than complex then this would allow an undesirable alternative $SO(3)$
invariant contraction of the indices in the Yukawa operator in
Eq.~(\ref{Yuk_op}), arising from the fact that for real triplets ${\bf 3
  \times 3}$ contains a singlet $\bf 1$ (whereas for complex representations
it does not), namely $\psi_{i}\psi_{i}^{c}\bar{\phi}^{j}\bar{\phi}^{j}H$
leading to a Yukawa matrix proportional to the unit matrix which would tend to destroy any hierarchies
in the Yukawa matrix. For real representations, there is no symmetry at the effective operator level that could forbid
such a trivial contraction of the two triplet fermion fields,
together with the trivial contraction of the two triplet flavon fields.
However in principle it is possible to appeal to the details
of the underlying theory responsible for generating this operator in order to forbid the trivial contraction,
which would involve a discussion of the heavy messenger states whose exchange generates the operator,
though we shall not pursue this possibility further here.

Alternatively, there is the possibility of obtaining a Yukawa operator
from a single triplet flavon field, $\phi^k$, using
$s_{ijk}\psi^{}_{i}\psi_{j}^{c}\phi^{}_{k}H$,
where $s_{ijk}$ is some tensor contraction. For example if the fields are triplets
of $SU(3)$ or $SO(3)$, then we can have $s_{ijk}=\epsilon_{ijk}$, the totally antisymmetric tensor.
However such operators cannot generate the diagonal elements of the Yukawa matrix.
For discrete groups such as $A_4$ the tensor $s_{ijk}$ can additionally be
chosen totally symmetric. Depending on the basis of the triplet
representation, it may however still be impossible to generate the diagonal elements
of the Yukawa matrix (see e.g. \cite{A4-LQ:King}). Therefore, in typical
models of this kind based on $A_4 \otimes SO(10)$,
the singlet contraction $\psi_{i}\psi_{i}^{c}\eta H$,
where $\eta$ is an $A_4$ singlet,
are typically invoked to obtain the diagonal elements, and
the problem of achieving a hierarchical Yukawa matrix is challenging \cite{A4-LQ:Morisi}.

The above considerations seem to point towards a $G_{f}\otimes
SO(10)$ theory, where $G_{f}$ is $SU(3)_{f}$
\cite{SU(3)-LQ:Ross} or $\Delta(27)$
\cite{delta27-LQ:King} since these groups contain
complex triplet representations. However there is one undesirable
feature of such theories, namely that the top Yukawa coupling must
necessarily originate from a higher order operator involving two
flavon fields, as in Eqs.~(\ref{Yuk_op}) and (\ref{Yuk}). This
implies that the flavon VEV must be quite close in magnitude to
the mass scale of the operator leading to a convergence problem,
since the effective top quark Yukawa coupling $h_t$ is typically
about 0.5 at the GUT scale in SUSY models, which implies that
$h_t=yV^2 /M^2 \approx 0.5$, or $V /M \approx 0.7y^{-1/2}$. We
emphasise that this problem is unavoidable in $G_{f}\otimes
SO(10)$ models unless the Higgs fields carry $G_{f}$ quantum
numbers which would allow renormalisable Yukawa couplings, but
such theories face the challenge of flavour changing neutral
currents.

\begin{figure}
\begin{center}
\includegraphics[width=12.5cm]{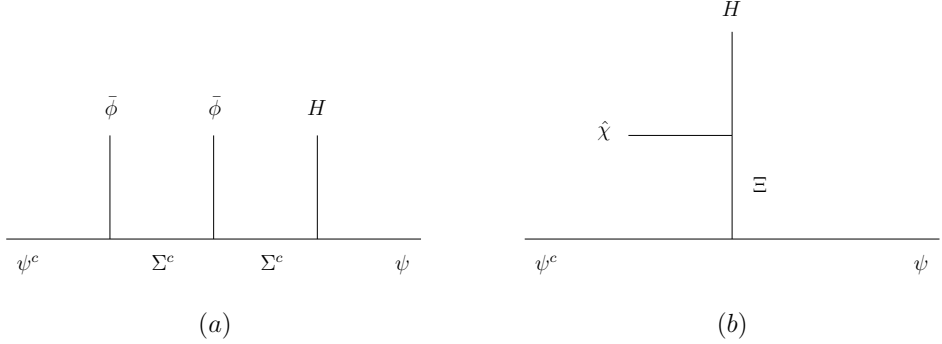}
\end{center}
\caption{Diagrams responsible for the Yukawa couplings. With
  anti-triplet flavons, case~$(a)$, the messengers $\Sigma^c$ are $SU(2)_L$
  singlets; with   (anti-)sextet flavons, case~($b$), the messengers are
  $SU(2)_L$   doublets.  \label{diagrams}}
\end{figure}

There is a further problem arising from anti-triplet flavons in the
explicit $SU(3)_{f}$ \cite{SU(3)-LQ:Ross} or $\Delta(27)$
\cite{delta27-LQ:King} models. It is related to the origin
of the top and bottom quark Yukawa coupling due to the diagram in
Fig.~\ref{diagrams}($a$) which is mediated by $SU(2)_L$ singlet
messengers $\Sigma^c$ which are supposed to be lighter than
$SU(2)_L$ doublet messengers and hence give the dominant contribution to the
effective Yukawa operator. Fig.~\ref{diagrams}($a$) involves anti-triplet
flavons $\bar \phi$ and is mediated by heavy fermionic messengers $\Sigma^c$
of two types namely $\Sigma_U^c$ and $\Sigma_D^c$
with mass $M_U$ and $M_D$, corresponding to the diagrams involving
up and down quark CP conjugated right-handed fields $U^c$ and $D^c$.
Similar diagrams for the first and second family quarks require
a messenger hierarchy in the up and down Yukawa sectors, $M_U/M_D \approx 3$,
which is necessary to account for the two different expansion parameters in the
two sectors, $\epsilon_u = v/M_U \approx 0.05$ and $\epsilon_d =
v/M_D \approx 0.15$, where $v$ refers to the first and second
flavons. This must be corrected for in the third family to achieve
$h_t\approx h_b \approx 0.5$.
This is done in a rather cumbersome way by assigning the
third family flavon $\phi_3$ to be a combination of a singlet plus
a triplet of $SU(2)_R$, and adjusting the flavon VEVs to lead to $h_t\approx h_b \approx 0.5$
\cite{SU(3)-LQ:Ross,delta27-LQ:King},
where the larger values of these Yukawa couplings is achieved by assuming
larger third family flavon VEVs $V>v$, as discussed previously.

It is possible to alleviate both the above problems by invoking (anti-)sextet
flavon fields as the origin of the third family Yukawa couplings. For example, introducing a two index
anti-sextet of $SU(3)_f$, $\hat{\chi}^{ij}$, which is symmetric under interchange of $i\leftrightarrow j$,
the lowest order Yukawa operators become,
\be
\frac{y}{M}\hat{\chi}^{ij}\psi^{}_{i}\psi_{j}^{c}H \ .
\label{Yuk_op2}
\ee
If the anti-sextet flavon $\hat{\chi}$ has a VEV aligned along
the 33 direction $\langle \hat{\chi}^{ij} \rangle = V\delta_{i3}\delta_{j3}$,
then this operator would imply a Yukawa matrix of the form
\begin{equation}
Y = \left(
\begin{array}{ccc}
0 & 0 & 0 \\
0 & 0 & 0 \\
0 & 0 & \frac{yV}{M}
\end{array}
\right) \ , \label{Yuk2}
\end{equation}
with the top quark Yukawa coupling of 0.5 implying $V /M \approx 0.5y^{-1}$,
which has somewhat improved convergence properties. In addition, unlike in
\cite{SU(3)-LQ:Ross,delta27-LQ:King}, it is
sufficient to assume that the third family flavon $\hat{\chi}$ is a singlet of
$SU(2)_R$ since the messengers will be qualitatively different in the case of
the third family due to the fact that $\hat{\chi}$ is a flavon anti-sextet
rather than an anti-triplet, as we now discuss.

According to the above discussion it is plausible to assume
that the operators responsible for the third family Yukawa couplings
involve (anti-)sextet flavons and are generated from diagrams
involving messengers $\Xi$ which carry Higgs quantum numbers as well as being
(anti-)sextets, see Fig.~\ref{diagrams}($b$).
Such Higgs messengers would imply similar mass
scales for the top and bottom Yukawa operators, naturally
and elegantly leading to top-bottom unification, with the
(anti-)sextet flavons being $SO(10)$ singlets.
The operators for the first and second family quarks could then continue to
involve anti-triplet flavons, with the relevant operators, given
in \cite{SU(3)-LQ:Ross}, arising from
diagrams involving heavy fermionic messengers $\Sigma^c$ of
mass $M_U$ and $M_D$, similar to Fig.~\ref{diagrams}($a$).

\begin{figure}
\begin{center}
\includegraphics[width=12.5cm]{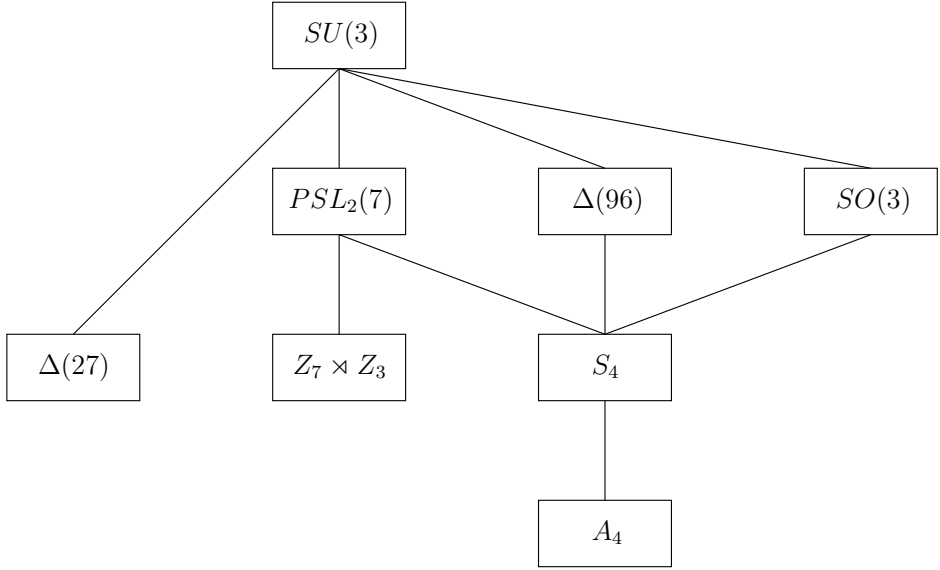}
\end{center}
\caption{$PSL_2(7)=\Sigma(168)$ and its relation to some other subgroups of
  $SU(3)$. A   line connecting two groups indicates that the smaller is a
  subgroup of the bigger  one. \label{family}} 
\end{figure}

The above discussion provides a good motivation for considering flavon
(anti-)sextets as the origin of the third 
family Yukawa couplings in $G_{f}\otimes SO(10)$ models. Although this is possible in the case
where the family group $G_{f}$ is $SU(3)_{f}$ \cite{SU(3)-LQ:Ross}, it is not possible for the
$\Delta(27)$ \cite{delta27-LQ:King} models for the simple reason that $\Delta(27)$ does
not admit (anti-)sextet representations. On the other hand, discrete family symmetry groups have been shown to be preferred
to continuous family symmetry groups such as $SU(3)_{f}$. The reason is that,
starting from a discrete symmetry group which contains a finite
number of group elements, it is easier to achieve
vacuum alignments which preserve the subgroup elements $S,U$ required for
tri-bimaximal neutrino mixing, rather than starting from a continuous group
which contains an infinite number of group elements. The question then
naturally arises of whether there is a discrete subgroup of $SU(3)_{f}$ which
admits triplet representations, to which we can assign the three families of quarks and leptons,
together with (anti-)sextet representations suitable for accommodating the flavons responsible for the third family Yukawa couplings.
We further require the triplet representations to be complex, to forbid the
singlet contraction $\psi_{i}\psi_{i}^{c}\eta H$, but the (anti-)sextet representation is allowed to be real or complex.
The smallest discrete group consistent with these requirements is ${PSL_2(7)}$, which is the
projective special linear group of two dimensional matrices over the finite Galois field of seven elements.
${PSL_2(7)}$ contains 168 elements and is also known as $\Sigma(168)$ \cite{FFK}.
For practical purposes, the most important features of ${PSL_2(7)}$ are that it admits complex triplet
and anti-triplet representations as well as a (real) sextet. ${PSL_2(7)}$ is a subgroup
of $SU(3)$, and contains $S_4$ as a maximal subgroup. By contrast, $\Delta(27)$ is also a subgroup
of $SU(3)$ but is not a subgroup of ${PSL_2(7)}$, and does not contain $S_4$ as a subgroup.
The relationship of ${PSL_2(7)}$ to some other family symmetries that have been used
in the literature is discussed in \cite{Z7Z3-LQ,PSLgroup,discAnom}
and depicted in Fig.~\ref{family}. From a mathematical point of view
${PSL_2(7)}$ is the unique finite group that is both {\it simple}
and contains complex triplet representations.\footnote{There exists another
simple group with triplet representations, the icosahedral group
$A_5$. Being a subgroup of $SO(3)$, its two triplets are real which gives rise
to the trivial bilinear invariant. Moreover, $S_4$ is not a subgroup of $A_5$
so that tri-bimaximal mixing cannot be obtained naturally; instead the solar
mixing may be related to the golden ratio~\cite{A5}.}

In this paper we shall develop the representation theory of
$PSL_2(7)=\Sigma(168)$ 
for triplets and sextets in a convenient basis suitable for applications
of ${PSL_2(7)}$ as a family symmetry capable of describing quark and lepton
masses and mixing in the framework of $SO(10)$ type unified models.
The representation theory of ${PSL_2(7)}$ has been developed in \cite{PSLgroup}
in terms of a standard presentation involving two generators $A,B$.
However, for physical applications, it is desirable to be able to
relate the representations of ${PSL_2(7)}$ to those of
$S_4$ in a basis in which the generators of $S_4$ are denoted as
$S,T,U$ where $S,U$ are the symmetry transformations of the
tri-bimaximal neutrino mass matrix in the diagonal charged lepton mass basis.
We shall show how the triplet representation given in terms of the
standard generators $A,B$ in \cite{PSLgroup} may be
related to four ${PSL_2(7)}$ generators which we shall denote as
$S,T,U,V$ where $S,T,U$ are the triplet generators of
its maximal subgroup $S_4$ in the physical basis described above.
In such a basis the subgroup structure ${PSL_2(7)} \supset S_4 \supset A_4$
just corresponds to the respective generators being $S,T,U,V \supset S,T,U \supset S,T$.
We shall also construct explicitly the four generators of
${PSL_2(7)}$ in a real sextet representation
${\mathcal S},{\mathcal T},{\mathcal U},{\mathcal V}$ which we relate to the
 couplings of the neutrino operators and third family
charged fermions. We identify those sextet vacua which preserve
the generators ${\mathcal S},{\mathcal U}$ as required for tri-bimaximal
neutrino mixing, or alternatively break all the generators and give rise to a
large third family charged fermion Yukawa coupling, such as the top quark
Yukawa coupling.

The layout of the remainder of the paper is as follows.
In section \ref{TB} we discuss the symmetries of the TB neutrino mass matrix in the
diagonal charged lepton basis and shall discover from this bottom-up procedure
the matrix transformations $S,U$ which leave the TB neutrino mass matrix invariant and
which correspond to two of the generators of $S_4$ in this physical basis.
In section \ref{triplet} we shall transform the triplet representations of
${PSL_2(7)}$ in the $A,B$ presentation to the desired $S,T,U,V$ form.
In section \ref{sextet} we shall repeat an analogous exercise for the sextet,
by defining a convenient real sextet basis ${\mathcal S},{\mathcal
  T},{\mathcal U},{\mathcal V}$ in terms of which the couplings involving
a sextet flavon field may be simply expressed. Section \ref{towards} outlines
the central steps in building a realistic $PSL_2(7)$ flavour model, whose
construction, including a detailed discussion of the dynamics, will be
presented in a separate publication~\cite{future}.
We conclude our presentation of a
$PSL_2(7)$ family symmetry in the framework of $SO(10)$ type GUT models in
section \ref{conclusion}.




\section{The symmetries of tri-bimaximal neutrino mixing \label{TB}}
\cleqn

The typical Lagrangian (or superpotential) of interest generically consists of
two parts, the Yukawa sector and the Majorana sector.
The effective Yukawa sector is of the form,
\be
{\mathcal L}^{Yuk} = \psi_{i}Y^{Yuk}_{ij}\psi_{j}^{c}H \ ,
\label{opYuk}
\ee
while the effective Majorana sector is of the form
\be
{\mathcal L}^{Maj} = \psi_{i} Y^{Maj}_{ij}\psi_{j}HH \ ,
\label{opMaj}
\ee
where $Y^{Yuk}_{ij}$ and $Y^{Maj}_{ij}$ are Yukawa and Majorana couplings, respectively,
while $H$ are Higgs fields. When the Higgs develop their VEVs,
and $\psi$ are identified with left-handed lepton fields $L$, while $\psi^{c}$ are
identified with charged conjugated right-handed charged leptons such as $e^c$,
the effective Yukawa operators lead to the charged lepton mass matrix
$M^e_{ij}\propto Y^e_{ij}$, while the effective Majorana operators lead to a
neutrino Majorana mass matrix $M^{\nu}_{ij}\propto Y^{\nu}_{ij}$. 

In the flavour basis, in which the charged lepton mass matrix is
diagonal and the TB mixing arises from the neutrino sector, the effective neutrino
mass matrix, denoted by ${M^{\nu}_{TB}}$, may be diagonalised as,
\begin{equation}
{M^{\nu}_{\mbox{\scriptsize diag}} } = U_{TB}^{T}
\,{M^{\nu}_{TB}} \,U_{TB}=\mathrm{Diag}\,(m_{1}, \; m_{2}, \; m_{3}) \; .
\end{equation}
Given $U_{TB}$, this
enables ${M^{\nu}_{TB}}$ to be determined in terms of the three
neutrino masses,
\begin{equation}\label{eq:csd-tbm0}
{M^{\nu}_{TB}}= \frac{m_{1}}{6} A + \frac{m_{2}}{3} B + \frac{m_{3}}{2} C \; ,
\end{equation}
where the three matrices are
\begin{equation}\label{eq:csd-tbmdec0}
A =
\left(\begin{array}{ccc}
4 & -2 & -2 \\
-2 & 1 & 1 \\
-2 & 1 & 1
\end{array}\right), \quad
B =
\left(\begin{array}{ccc}
1 & 1 & 1 \\
1 & 1 & 1 \\
1 & 1 & 1
\end{array}\right), \quad
C = \left(\begin{array}{ccc}
0 & 0 & 0 \\
0 & 1 & -1 \\
0 & -1 & 1
\end{array}\right) \; .
\end{equation}
From above we may write ${M^{\nu}_{TB}}$ as the symmetric matrix,
\begin{equation}\label{eq:csd-tbm2}
{M^{\nu}_{TB}}~=~
\left(\begin{array}{ccc}
a & b & c \\
b & d & e \\
c & e & f
\end{array}\right),
\end{equation}
where,
\bea
a= \frac{2m_1}{3}+\frac{m_2}{3} \ , \quad
b = c = -\frac{m_1}{3}+\frac{m_2}{3} \ , \quad
d = f = a+b-e = \frac{m_1}{6}+\frac{m_2}{3} +\frac{m_3}{2} \ .\label{abc}
\eea
In particular $b=c$ and $d=f$ and $a+b= d+e$
are the characteristic signatures of the TB neutrino mass matrix in the flavour basis.

It seems paradoxical that, in this basis, ${\mathcal L}^{Yuk}$ seems to obey a
completely different symmetry as compared to that of ${\mathcal
  L}^{Maj}$. While ${\mathcal L}^{Yuk}$ only respects a trivial phase
symmetry corresponding to a diagonal phase transformation on the fields, which leaves
the diagonal mass matrix $M^e$ invariant, ${\mathcal L}^{Maj}$ respects
a more subtle discrete symmetry which leaves ${M^{\nu}_{TB}}$ invariant,
even through the same fields, namely the lepton doublets $L$, are common to
both. How can this be reconciled? If the underlying couplings $Y^{Yuk}_{ij}$
and $Y^{Maj}_{ij}$ are simply numbers, then there is no way that different
parts of the same Lagrangian involving the same fields could have different
symmetries: it would not make any sense since any symmetry transformation
$N$ on the lepton doublets $L\rightarrow N L$ must transform both ${\mathcal
 L}^{Yuk}$ and ${\mathcal L}^{Maj}$.
The resolution to this problem is intrinsically related to the origin of the
Yukawa couplings. If the Yukawa couplings are generated by the VEVs of flavon
fields, as is generically the case for models which respect a family symmetry,
then it is possible to have different symmetries in the Yukawa and Majorana
sectors, providing different flavons appear, or are dominant, in the two
sectors. The idea is that the complete high energy theory Lagrangian,
including both ${\mathcal L}^{Yuk}$ and ${\mathcal L}^{Maj}$, would both
respect some family symmetry $G_{f}$ due to the presence of flavons of two
types $\phi^{Yuk}$ and $\phi^{Maj}$, where $\phi^{Yuk}$  appears in ${\mathcal
  L}^{Yuk}$ but not in ${\mathcal L}^{Maj}$, where only $\phi^{Maj}$ appears.
In other words the effective Yukawa sector is supposed to originate from terms
of the form, 
\be
{\mathcal L}^{Yuk} = \psi^{}_{i}\phi_{ij}^{Yuk}\psi_{j}^{c}H \ ,
\label{opYuk-1}
\ee
while the effective Majorana sector comes from terms like
\be
{\mathcal L}^{Maj} = \psi^{}_{i}\phi_{ij}^{Maj}\psi^{}_{j}HH \ ,
\label{opMaj-1}
\ee
where both terms are invariant under $G_{f}$, with the indices $i,j$ now being
regarded as indices of $G_{f}$. Generically $\phi^{Yuk}$ and $\phi^{Maj}$ may
represent either a single flavon or a polynomial of flavons of a particular
type. When the flavons develop VEVs the family symmetry is spontaneously
broken, so that the full family symmetry is not apparent in either of the
low energy Lagrangians ${\mathcal L}^{Yuk}$ or ${\mathcal L}^{Maj}$.
However, although both flavon types break the family symmetry, if $\phi^{Yuk}$
and $\phi^{Maj}$ preserve two different subgroups of $G_{f}$, say $G^{Yuk}$
and $G^{Maj}$, then this will result in the two low energy Lagrangian terms
${\mathcal L}^{Yuk}$ and ${\mathcal L}^{Maj}$ respecting different
symmetries $G^{Yuk}$ and $G^{Maj}$, as observed.

In the case of TB mixing, we have seen that, in the diagonal charged lepton basis, after the Higgs develop their VEVs,
the Majorana Lagrangian is
\be
{\mathcal L}^{Maj}_{TB} = \psi^{}_{i}({M^{\nu}_{TB}})_{ij} \psi^{}_{j} \ ,
\label{opMajTB}
\ee
where ${M^{\nu}_{TB}}$ is given in Eq.~(\ref{eq:csd-tbm0}). According to the discussion of the preceding paragraph,
$({M^{\nu}_{TB}})_{ij}\propto \langle \phi^{Maj}_{ij}\rangle$,
so it makes sense to identify the symmetry $G^{Maj}_{TB}$ respected by
${\mathcal L}^{Maj}_{TB}$ in Eq.~(\ref{opMajTB}), independently of the
symmetry of the charged lepton sector. We want to find the most general
symmetry transformation $N$, corresponding to $\psi \rightarrow N\psi$, which
leaves ${\mathcal L}^{Maj}_{TB}$ in Eq.~(\ref{opMajTB}) invariant. This will
subsequently be identified with the symmetry respected by $\langle
\phi^{Maj}_{ij}\rangle$, which will be a subgroup of the full family symmetry
$G_f$. Assuming that $N$ is a unitary matrix, the symmetry condition is given by,
\be
{M^{\nu}_{TB}}\, N= \bar N \, {M^{\nu}_{TB}} \ ,
\label{S}
\ee
with $\bar N$ denoting the complex conjugate of $N$.
Since the symmetry must be independent of the neutrino mass eigenvalues,
Eq.~(\ref{S}) implies, from Eqs.~(\ref{eq:csd-tbm0}) and
(\ref{eq:csd-tbmdec0}), three independent conditions,
\be
A \,N~=~ \bar N A \ , \qquad
B \,N~=~ \bar N B \ , \qquad
C \,N~=~ \bar N C \ .
\label{S3}
\ee
Starting with the most general complex ($3 \times 3$) matrix, these conditions
readily lead to the real matrix
\be
N ~=~ \begin{pmatrix}
n_1 & n_2 & n_2 \\
n_2 & n_3 & n_1+n_2-n_3 \\
n_2 & n_1+n_2-n_3 & n_3
\end{pmatrix} \ .
\ee
Since $N$ appears quadratically in the conditions of Eq.~(\ref{S3}), it is
only determined up to an overall sign. Assuming that $N$ has determinant $+1$,
for definiteness, the requirement of unitarity (orthogonality) finally yields
the most general  symmetry of the TB neutrino mass matrix
$M^{\nu}_{TB}$ as the group obtained from the generators $S,U$, given by,
\begin{equation}
S = \frac{1}{3} \left(\begin{array}{ccc}
-1& 2  & 2  \\
2  & -1  & 2 \\
2 & 2 & -1
\end{array}\right), \qquad
U = - \left( \begin{array}{ccc}
1 & 0 & 0 \\
0 & 0 & 1 \\
0 & 1 & 0
\end{array}\right) \ .\label{SUgens}
\end{equation}
It is easy to show that this group has only four elements; excluding the
identity, all are of order two. Hence we have discovered Klein's four-group
$Z_2 \times Z_2$ as the symmetry of the TB neutrino mass matrix.

By contrast, in the diagonal charged lepton mass basis,
${\mathcal L}^{Yuk}$ satisfies a rather trivial diagonal phase symmetry. The
first possible choice of phases which reversely enforce a diagonal charged
lepton mass matrix corresponds to powers of the diagonal symmetry
generator~$T$, given by
\begin{equation}
T = \left( \begin{array}{ccc}
1 & 0 & 0 \\
0 & \omega^{2} & 0 \\
0 & 0 & \omega
\end{array}\right) \ ,\label{Tgen}
\end{equation}
where $\omega = e^{2\pi i/3}$.

As we will prove in the following section, the matrices $S,T,U$ form the
generators of the group $S_4$ in the triplet representation, and thus,
according to the previous discussion, $S_4$ or any group containing $S_4$
becomes an attractive candidate family symmetry $G_f$.




\section{The ${\boldsymbol{PSL_2(7)}}$ triplet in the physical ${\boldsymbol{S_4}}$ basis\label{triplet}}
\cleqn

As outlined in the previous section, a natural
family symmetry should contain the non-Abelian finite group $S_4$,
the permutation group on four letters. As both triplet
representations of $S_4$ itself are real one can form the trivial
invariant from ${\bf 3 \times 3}$ and ${\bf 3' \times 3'}$.
Therefore the hierarchy of the fermion masses cannot be explained
(without fine-tuning) if we want to work in a unified framework
where the quarks and leptons all transform as triplets ${\bf 3}$ or
alternatively ${\bf 3'}$ under the family group $S_4$.

Requiring complex triplet and (real or complex) sextet
representations, the smallest finite groups that have $S_4$ as a
subgroup are $\Delta(96)$ and $PSL_2(7)=\Sigma(168)$. The former belongs to
the series of groups $\Delta(6n^2)$ and has one sextet, six
triplet, one doublet and two singlet representations. It is
interesting to note that two of the triplets are real while the
other four are complex. The irreducible representations of
$PSL_2(7)$ on the other hand are
$$
PSL_2(7) : ~~ {\bf 1,~3,~\overline 3,~6,~7,~8} \ .
$$
As there are only two complex triplet representations (${\bf 3,\overline 3}$),
the group $PSL_2(7)$ appears more attractive from the model building point of
view. Furthermore, as already mentioned in the introduction, it is the unique
{\it simple} finite group with complex triplet representations which makes
$PSL_2(7)$ appealing also from the mathematical perspective.

In the following we want to pave the way for building flavour models based on
the group $PSL_2(7)$. The main idea is to use the crucial ingredients of the
known $A_4$ and $S_4$ models and generalise these in the context of a
$PSL_2(7)$ symmetric theory. In particular, we have to find a formulation in
which the groups $A_4$, $S_4$, and $PSL_2(7)$ can be easily compared. This
involves two separate questions:
\begin{enumerate}
\item What is the structure of the embedding $A_4 \subset S_4 \subset
  PSL_2(7)$ in terms of generators?
\item What is the most convenient basis for the matrix representations of
  these generators?
\end{enumerate}



\subsection{${\bs{A_4 \subset S_4}}$}

Before discussing the group $PSL_2(7)$, let us illustrate the situation for
the less complicated case where $A_4$ is embedded into $S_4$. These groups
belong to the series of groups $\Delta(3n^2)$ and $\Delta(6n^2)$,
respectively. With $n=2$, we have
\begin{eqnarray}
A_4  &=& \Delta(12)~=~ (\underbrace{Z_2}_{c} \times \underbrace{Z_2}_{d})
\rtimes \underbrace{Z_3}_{a} \ ,\label{a4}\\[2mm]
S_4 &=&\Delta(24)~=~ (\underbrace{Z_2}_{c} \times \underbrace{Z_2}_{d})
\rtimes (\underbrace{Z_3}_{a} \rtimes \underbrace{Z_2}_{b}) \ ,\label{s4}
\end{eqnarray}
where the generators $c,d,a$, and $b$ of the various cyclic groups have to
satisfy the following relations
\begin{eqnarray}
&c^2 = d^2 = a^3 = 1\,,\quad cd=dc\,,\quad aca^2 = cd\,,\quad ada^2 = c \ ,\label{A4S4}\\
&b^2 = (ab)^2 = 1\,,\quad bcb=d \ .\label{onlyS4}
\end{eqnarray}
This presentation has the advantage that the embedding of $A_4$ into $S_4$ is
explicitly given: adding a generator $b$ which satisfies the relations of
Eq.~(\ref{onlyS4}) to the group $A_4$ leads to~$S_4$. However, in practice,
model builders adopt a slightly different convention for the presentation of
$A_4$. Denoting the generators by $s$ and $t$, this alternative presentation
takes the compact form
\begin{equation}
A_4: ~~ < s,t \,|\, s^2 = t^3 = (st)^3 = 1 > \ .\label{A4pres}
\end{equation}
The first observation is that it is sufficient to work with only two instead
of three generators. This can be easily understood by noting that the
generator $c$ is nothing but $ada^2$, as required by the last relation in
Eq.~(\ref{A4S4}). Therefore $c$ can be replace by $ada^2$ and becomes
redundant in the presentation. Having dropped the generator $c$, we need to
identify $s$ and $t$ among the 12 group elements of $A_4$ given in terms of
the original generators $d$ and $a$. Since $s$ and $d$ are both elements of
order two, while $t$ and $a$ are elements of order three, it is suggestive to
set
\begin{equation}
s ~=~ d \,, \qquad t ~=~ a \ .\label{A4choice}
\end{equation}
One can show that this identification is consistent by inserting it into
Eq.~(\ref{A4S4}) and rewriting the result to obtain the presentation of
Eq.~(\ref{A4pres}). However, other identifications are equally possible. For
instance, we could have chosen $s'=c=ada^2$ and $t'=a$ which is related to
Eq.~(\ref{A4choice}) by the similarity transformation $s'=asa^2$ and
$t'=ata^2$. In the following, we disregard such an ambiguity and stick to the
identification given in Eq.~(\ref{A4choice}).

When applying a non-Abelian discrete family symmetry, one is mainly interested
in the matrix representations for the triplets. In the presentation of $S_4$
with $c,d,a$, and $b$ they are given in~\cite{deltagroups}:\footnote{A
  different presentation for
  $S_4$ is given in~\cite{S4-LQ:Hagedorn} by $<\tilde A,\tilde
  B\,|\,\tilde A^4 = \tilde B^3 = (\tilde A\tilde B^2)^2 = 1>$. Using the
  explicit matrix representation for the triplets, the generators $\tilde A$ and
  $\tilde B$ are related to $a,b,c,d$ through $\tilde A= bda$ and  $\tilde
  B=a^2$. It is straightforward to show that this presentations is
  equivalent to the one shown in Eqs.~(\ref{A4S4}) and~(\ref{onlyS4}).}
\begin{eqnarray}
c^{[{\bf 3}]}\,=\begin{pmatrix}  -1 & 0 & 0 \\ 0 & -1 & 0 \\ 0& 0 & 1 \end{pmatrix} ,&
~~~
d^{[{\bf 3}]}\,=\begin{pmatrix}  1 & 0 & 0 \\ 0 & -1 & 0 \\ 0& 0 &
  -1 \end{pmatrix} ,&
~~~
a^{[{\bf 3}]}\,=\begin{pmatrix} 0  & 1 & 0 \\ 0 & 0 & 1 \\ 1 & 0 & 0 \end{pmatrix} ,\label{datrip}\\[3mm]
&~\,~b^{[{\bf 3}]}\,=\,\pm \begin{pmatrix}  0 & 0 & 1 \\ 0 & 1 & 0 \\ 1& 0 & 0 \end{pmatrix} .\,\label{btrip}
\end{eqnarray}
Notice that $b^{[{\bf 3}]}$ has two different matrix representations
corresponding to the
two triplets of $S_4$. Of course, these explicit matrices are defined in a
particular basis for the triplets. Even though $A_4$ models were originally
formulated using the basis of Eq.~(\ref{datrip}), a  different basis has
proved more useful in physical applications. Suppressing the redundant
generator $c^{[{\bf 3}]}$, we define for the triplet representations
\begin{eqnarray}
S
= w \,d^{[{\bf 3}]}\, w^\dagger , \qquad
T
= w \,a^{[{\bf 3}]}\, w^\dagger , \qquad
U'= w \,b^{[{\bf 3}]}\, w^\dagger , \label{STU'trafo}
\end{eqnarray}
with the unitary matrix $w$ given by
\begin{equation}
w~=\frac{1}{\sqrt{3}} \begin{pmatrix} 1&1&1\\1 &\omega&\omega^2 \\ 1
  &\omega^2&\omega \end{pmatrix} \ , \qquad \omega = e^{2 \pi i/3} \ .
\end{equation}
We then obtain the explicit $S_4$ generators in the new basis
\begin{equation}
S~= \frac{1}{3} \begin{pmatrix} -1&2&2\\2 &-1&2 \\ 2&2&-1 \end{pmatrix}
 , ~\quad
T~= \begin{pmatrix} 1&0&0\\0 &\omega^2&0 \\ 0&0&\omega \end{pmatrix}  ,
~\quad
U'~= \, \pm \begin{pmatrix} 1&0&0\\0 &0&\omega^2 \\ 0&\omega&0 \end{pmatrix} \ .
\end{equation}
Now $S$ and $T$ are in the conventional form suitable for $A_4$ model
building. $S_4$ is obtained by simply adding the generator $U'$ or,
alternatively, a  product of $U'$ with any group element of $A_4$, e.g.
\begin{equation}
U ~=~T^2 \, U' ~= \,\pm \begin{pmatrix} 1&0&0\\0 &0&1 \\ 0&1&0 \end{pmatrix}\ .\label{U'toU}
\end{equation}
With the matrices $S$, $T$, and $U$ we have recovered the generators of $S_4$
for the triplet representations found by Lam in~\cite{Lam}, with the
identifications $S = G_2$, $T=(F_1)^2$,
and $U=G_3$. Note that they are identical to the matrices in
Eqs.~(\ref{SUgens}) and (\ref{Tgen}).



\subsection{${\bs{S_4 \subset PSL_2(7)}}$}

Having identified the embedding of $A_4$ in $S_4$ in terms of the physically
interesting generators $S,T$, and $U$, we want to proceed with the discussion
of the group $PSL_2(7)$. Details of its mathematical structure -- e.g. character
table, irreducible representations, Kronecker products -- can be found
in~\cite{PSLgroup}. There, the presentation of $PSL_2(7)$ is given in the
compact form
\begin{equation}
PSL_2(7): ~~ <A,B\,|\,A^2 \,=\, B^3 \,=\, (AB)^7 \,=\, (A^{-1}B^{-1}AB)^4 \,=\, 1> \ .\label{pslpres}
\end{equation}
In order to extract a set of 24 elements that constitutes $S_4$ within the 168
elements of $PSL_2(7)$, we first identify the order three generators $a$ and
$B$. This choice can be made without loss of generality since all order three
elements belong to one conjugacy class and are therefore related to each other
by a similarity transformation. With this choice, the $S_4$ generators $d,a,b$
are related to the $PSL_2(7)$ generators $A,B$ by
\begin{equation}
d\,=\,[(AB)^2 B (AB)]^2\,,\qquad
a\,=\,B\,,\qquad
b\,=\,B^2(AB)(AB^2)^2 [(AB)^2 B (AB)]^2   \,.\label{s4inpsl}
\end{equation}
Relying only on the $PSL_2(7)$ presentation of Eq.~(\ref{pslpres}), these
identifications can be shown to satisfy the relations in Eqs.~(\ref{A4S4}) and
(\ref{onlyS4}) regardless of a concrete matrix representation. The complete
group $PSL_2(7)$ can be obtained by adding a generator $v$ which, for later
convenience, we choose to be
\begin{equation}
v~=~ (b d a b) \, A \, (b d a b)^{-1}  \ .
\end{equation}
It is worth noting that, contrary to $A_4$ and $S_4$ which are given in
Eqs.~(\ref{a4}) and (\ref{s4}), $PSL_2(7)$ cannot be written as a semidirect
product of $Z_n$ symmetries because it is a simple group. Nonetheless,
$PSL_2(7)$ can be obtained from one order three and three order two generators:
\begin{equation}
a^3\,=\,1 \ ,\qquad d^2\,=\,b^2\,=\,v^2 \,=\,1 \ .
\end{equation}
Here, we refrain from presenting the other conditions that these four generators
must satisfy since they are rather involved and do not provide any new insight
into the group structure.

As mentioned above, we are mostly interested in triplet representations.
The $(3\times 3)$ matrices $A^{[{\bf 3}]}$ and $B^{[{\bf 3}]}$ were
derived in~\cite{PSLgroup}. Defining $\eta = e^{2 \pi i/7}$, they take
the form
\begin{eqnarray}
A^{[\bf 3]}_{}&=&\frac{i}{\sqrt{7}}\begin{pmatrix} \eta^2-\eta^5 &\eta-\eta^6&
  \eta^4-\eta^3 \\ \eta-\eta^6&\eta^4-\eta^3&\eta^2-\eta^5 \\
  \eta^4-\eta^3&\eta^2-\eta^5&\eta-\eta^6\end{pmatrix} \ , \\[2mm]
B^{[\bf 3]}_{}&=&\frac{i}{\sqrt{7}}\begin{pmatrix} \eta^3-\eta^6 &\eta^3-\eta&
  \eta-1 \\ \eta^2-1&\eta^6-\eta^5&\eta^6-\eta^2\\
  \eta^5-\eta^4&\eta^4-1&\eta^5-\eta^3\end{pmatrix} \ .
\end{eqnarray}
Plugging these expressions into
Eq.~(\ref{s4inpsl}), we see that the resulting matrices for the $S_4$
generators are different from those presented in Eqs.~(\ref{datrip})
and~(\ref{btrip}). We must therefore find a similarity transformation that
relates both bases. This is facilitated by the observation that $a^{[{\bf
    3}]}$ is obtained from $B^{[{\bf 3}]}$ by conjugation
\begin{equation}
a^{[{\bf 3}]} ~=~ w_1 \, B^{[{\bf 3}]} \, w_1^\dagger \ ,
\end{equation}
where
\begin{equation}
w_1~=~
B^{[{\bf 3}]} (A^{[{\bf 3}]} B^{[{\bf 3}]} )^2 B^{[{\bf 3}]} (A^{[{\bf 3}]}
B^{[{\bf 3}]}) A^{[{\bf 3}]} \ .
\end{equation}
Applying this transformation to the other two $S_4$ generators shows that we
need a second transformation $w_2$ to arrive at the matrices of
Eqs.~(\ref{datrip}) and~(\ref{btrip}). However, $w_2$ is constrained in that
it must not change $a^{[{\bf 3}]}$, that is $w_2 \,a^{[{\bf 3}]} = a^{[{\bf
    3}]}\, w_2$. This condition allows only matrices of the form
\begin{equation}
w_2~=~\begin{pmatrix} x&y&z \\ z&x&y \\ y&z&x    \end{pmatrix} .
\end{equation}
Demanding $w_2$ to be unitary, we find the solution
\begin{eqnarray}
x &=& \frac{1}{N}\,(-64 - 15 \eta - 56 \eta^3 - 46 \eta^4 + 5 \eta^5 - 27\eta^6) \ , \\
y &=& \frac{1}{N}\,(73 \eta + 156 \eta^2 + 46\eta^3+12\eta^4+137\eta^5 + 115 \eta^6)\ ,\\
z &=& \frac{1}{N}\,(15 + 34 \eta - 35 \eta^2 - 23 \eta^3 + 41 \eta^4 - 46 \eta^6)\ ,
\end{eqnarray}
with the normalisation factor
\begin{equation}
N\,=\, \left[ 28(-342125 - 349668 \eta + 283769 \eta^2 + 9406 \eta^3 -  501928
  \eta^4 + 287955 \eta^6)\right]^{1/3}  .
\end{equation}
The combined similarity transformation $w_2\,w_1$ leads to the following matrix
representation for the $PSL_2(7)$ triplet
\begin{eqnarray}
d^{[{\bf 3}]}&=&w_2\,w_1\left[(A^{[{\bf 3}]}B^{[{\bf 3}]})^2 B^{[{\bf 3}]}
  (A^{[{\bf 3}]}B^{[{\bf 3}]})\right]^2  \,w_1^\dagger \,w_2^\dagger ~
=\,\begin{pmatrix} 1&0&0\\0&-1&0 \\0&0&-1\end{pmatrix} , \\
a^{[{\bf 3}]}&=&w_2\,w_1\,B^{[{\bf 3}]} \,w_1^\dagger \,w_2^\dagger ~
=\,\begin{pmatrix} 0&1&0\\0&0&1 \\1&0&0\end{pmatrix} , \\[3mm]
b^{[{\bf 3}]}&=&w_2\,w_1\,{B^{[{\bf 3}]}}^2(A^{[{\bf 3}]}B^{[{\bf
    3}]})(A^{[{\bf 3}]}{B^{[{\bf 3}]}}^2)^2 \left[(A^{[{\bf 3}]}B^{[{\bf
      3}]})^2 B^{[{\bf 3}]} (A^{[{\bf 3}]}B^{[{\bf 3}]})\right]^2
\,w_1^\dagger\, w_2^\dagger  \notag \\[2mm]
&=& - \, \begin{pmatrix} 0&0&1\\0&1&0 \\1&0&0\end{pmatrix} , \label{b3sign}\\
v^{[{\bf 3}]}&=&w_2\,w_1 \, (b^{[{\bf 3}]}d^{[{\bf 3}]}a^{[{\bf 3}]}b^{[{\bf 3}]})
A^{[{\bf 3}]}  (b^{[{\bf 3}]}d^{[{\bf 3}]}a^{[{\bf 3}]}b^{[{\bf 3}]})^{-1}
\,w_1^\dagger \,w_2^\dagger ~
=\,-\,\frac{1}{2}
\begin{pmatrix}
0\,&b_7&b_7 \\
\bar b_7 &1&-1\\
\bar b_7 &-1&1 \end{pmatrix}\!\!. ~~~~~~
\end{eqnarray}
Here we adopt the notation of the ``Atlas of Finite Groups'' \cite{atlas}
which defines
\begin{equation}
b_7 \,=\,\frac{1}{2} (-1+i\sqrt{7}) \ , \qquad \bar b_7 \,=\,\frac{1}{2}
(-1-i\sqrt{7}) \ .
\end{equation}
Notice that the sign of the generator $b^{[{\bf 3}]}$ is fixed in the
$PSL_2(7)$ framework. Hence there is only one triplet representation and its
conjugate (which is obtained by complex conjugation of $v^{[{\bf 3}]}$).

Finally, we can perform the basis transformation of
Eqs.~(\ref{STU'trafo}-\ref{U'toU}) to arrive at the $S_4$
generators\footnote{As for $b^{[{\bf 3}]}$ in Eq.~(\ref{b3sign}), the sign of
  $U$ is determined such that $\mathrm{det}(U)=1$.}
$S,T,U$ plus the additional generator
\begin{equation}
V~=~w\,v^{[{\bf 3}]} \, w^\dagger ~=\, \frac{1}{3}\,
\begin{pmatrix}
1 & 2 c_7 & 2 c_7  \\
2 \bar c_7 & -2 & 1\\
2 \bar c_7 & 1 &  -2 \end{pmatrix} , ~~~~ \text{with} ~~
\left\{\begin{array}{l} c_7\,=\,\frac{1}{8}( 1+  3 \, i \sqrt{7}) \ ,\\[2mm]
\bar c_7\,=\,\frac{1}{8}( 1 -  3 \, i \sqrt{7}) \ ,
\end{array} \right.
\end{equation}
which is necessary to obtain the remaining elements of $PSL_2(7)$. With these
four ($3\times 3$) matrices, we have found the structure of the embedding $A_4
\subset S_4 \subset PSL_2(7)$ in terms of the group generators:
\begin{eqnarray*}
S,T ~~~~~~\,\,\; & \longrightarrow & ~\quad A_4 \ ,\\
S,T,U ~~~\:\: & \longrightarrow & ~\quad S_4 \ ,\\
S,T,U,V \; & \longrightarrow & ~ PSL_2(7)=\Sigma(168) \ .
\end{eqnarray*}
Furthermore, $S,T,U$ are given in a basis for the triplets that is most
suitable for model building purposes \cite{Lam}.




\section{The ${\bs{PSL_2(7)}}$ sextet in a convenient real basis \label{sextet}}
\cleqn

Just like in $SU(3)$, the $PSL_2(7)$ sextet
$\tilde\chi=(\tilde\chi^{}_1,\tilde\chi^{}_2,\tilde\chi^{}_3,\tilde\chi^{}_4,
\tilde\chi^{}_5,\tilde\chi^{}_6)$ is generated from the symmetric product of two
triplets $\psi=(\psi_1,\psi_2,\psi_3)$ and
$\psi'=(\psi'_1,\psi'_2,\psi'_3)$. With
\bea
&
\tilde\chi^{}_1~=~ \psi^{}_1\psi'_1 \, ,~\quad
\tilde\chi^{}_2~=~ \psi^{}_2\psi'_2 \, ,~\quad
\tilde\chi^{}_3~=~ \psi^{}_3\psi'_3 \, , \nonumber \\
&
\tilde\chi^{}_4= \frac{1}{\sqrt{2}} (  \psi^{}_1 \psi'_2 + \psi^{}_2 \psi'_1 ) \, ,\quad
\tilde\chi^{}_5=\frac{1}{\sqrt{2}}  (  \psi^{}_2 \psi'_3 + \psi^{}_3 \psi'_2 ) \, ,\quad
\tilde\chi^{}_6=\frac{1}{\sqrt{2}}  (  \psi^{}_3 \psi'_1 + \psi^{}_1 \psi'_3 ) \, ,~~\label{6from3b}
\eea
the corresponding $PSL_2(7)$ generators for the sextet are
\bea
\tilde{\mathcal S}&=&\frac{1}{9}
\begin{pmatrix}
1&4&4&-2 \sqrt{2} & 4\sqrt{2} &-2 \sqrt{2} \\
4&1&4&-2 \sqrt{2} & -2\sqrt{2} &4 \sqrt{2} \\
4&4&1&4 \sqrt{2} & -2\sqrt{2} &-2 \sqrt{2} \\
-2 \sqrt{2} & -2\sqrt{2} &4 \sqrt{2}&5&2&2 \\
4 \sqrt{2} & -2\sqrt{2} &-2 \sqrt{2}&2&5&2 \\
-2 \sqrt{2} & 4\sqrt{2} &-2 \sqrt{2} &2&2&5
\end{pmatrix} ,
\eea

\bea
\tilde{\mathcal T}~=~
\begin{pmatrix}
1&0&0&0&0&0 \\
0&\omega&0&0&0&0 \\
0&0&\omega^2&0&0&0 \\
0&0&0&\omega^2&0&0 \\
0&0&0&0&1&0 \\
0&0&0&0&0&\omega \\
\end{pmatrix} ,
&\quad&
\tilde{\mathcal U}~=~
\begin{pmatrix}
1&0&0&0&0&0 \\
0&0&1&0&0&0 \\
0&1&0&0&0&0 \\
0&0&0&0&0&1 \\
0&0&0&0&1&0 \\
0&0&0&1&0&0
\end{pmatrix} ,
\eea

\bea
\tilde{\mathcal V}&=&\frac{1}{9}
\begin{pmatrix}
 1&-4+c_7&-4+c_7&2 \sqrt{2} \,c_7 & \sqrt{2}(-4+c_7)& 2 \sqrt{2} \, c_7\!\!\\
-4+\bar c_7 &4&1& -4\sqrt{2} \, \bar c_7  &-2\sqrt{2}& 2\sqrt{2}\,\bar c_7\!\!\\
 -4+\bar c_7&1&4&2\sqrt{2}\,\bar c_7 &-2\sqrt{2}&-4\sqrt{2} \, \bar c_7\!\!\\
 2\sqrt{2}\,\bar c_7&-4\sqrt{2} \, c_7&2\sqrt{2}\, c_7&2&-2c_7&5\\
 \!\!\sqrt{2}(-4+\bar c_7)&- 2 \sqrt{2}&- 2 \sqrt{2}&-2\bar c_7&5 &-2 \bar c_7\\
  2\sqrt{2}\,\bar c_7&2 \sqrt{2} \, c_7&-4\sqrt{2} \,c_7&5&-2 c_7&2
\end{pmatrix}\!\! .~~~~~~~~
\eea
The tilde denotes that this basis for the sextet is
complex. Since, unlike in $SU(3)$, the sextet of $PSL_2(7)$ is a real
representation, it is possible to change to a basis in which all generators are
explicitly real. In such a real basis, the bilinear invariant ${\bf 6 \otimes
  6}$ is trivially obtained by adding the squares of all components. In the
complex basis, we have to find a matrix $\tilde{\m C}$ such that
$
\tilde\chi^T \, \tilde{\m C} \, \tilde\chi \,
$
is invariant under all four generators $\tilde{\m S}, \tilde{\m T}, \tilde{\m U},
\tilde{\m V}$. A straightforward calculation yields
\bea
\tilde{\m C}&=&\frac{1}{6}
\begin{pmatrix}
2i-\sqrt{7}\,\bar b_7 & 0&0&0&i\sqrt{2}\,\bar b_7&0\\
0&0&-i-\sqrt{7}\,\bar b_7 & -\sqrt{14}\, b_7 &0&0 \\
0 &-i-\sqrt{7}\, \bar b_7 & 0&0&0&-\sqrt{14}\, b_7 \\
0&-\sqrt{14}\,b_7&0&0&0&2i\bar b_7 \\
i\sqrt{2}\, \bar b_7 &0&0&0&-2i-2\sqrt{7}\,\bar b_7 &0\\
0&0&-\sqrt{14} \,b_7&2i\bar b_7 &0&0
\end{pmatrix} . \notag
\eea
Using this matrix together with Eq.~(\ref{6from3b}), we
can now easily obtain the invariant combination of ${\bf 3 \otimes 3 \otimes
  6}$. After rearrangement of the terms, we find the $PSL_2(7)$ invariant
\be
-\frac{1}{6} ~
\psi^T \,   \hat\chi \,  \psi' \ ,
\ee
where we have defined the symmetric {\it matrix} $\,\hat \chi_{ij}$ as
\be
 \hat\chi =
\begin{pmatrix}
\! (-2i+\sqrt{7}\,\bar b_7) \tilde\chi_1 -i  \sqrt{2} \,\bar b_7 \tilde\chi_5\!&\!
\sqrt{7} \,b_7 \tilde\chi_2 -i \sqrt{2}\, \bar b_7 \tilde\chi_6 \!&\!
 \sqrt{7} \,b_7 \tilde\chi_3 -i \sqrt{2}\, \bar b_7 \tilde\chi_4\! \\[2mm]
\!\sqrt{7} \,b_7 \tilde\chi_2 -i \sqrt{2}\, \bar b_7 \tilde\chi_6 \!&\!
(i+\sqrt{7}\,\bar b_7) \tilde\chi_3 +   \sqrt{14} \,b_7 \tilde\chi_4\!&\!
\sqrt{2} (i+\sqrt{7} \,\bar b_7) \tilde\chi_5 -i \bar b_7 \tilde\chi_1 \!\\[2mm]
\! \sqrt{7} \,b_7 \tilde\chi_3 -i \sqrt{2}\, \bar b_7 \tilde\chi_4 \!&\!
\sqrt{2} (i+\sqrt{7} \,\bar b_7) \tilde\chi_5 -i \bar b_7 \tilde\chi_1 \!&\!
(i+\sqrt{7}\,\bar b_7) \tilde\chi_2 +   \sqrt{14} \,b_7 \tilde\chi_6 \!
\end{pmatrix}\!.\label{yukmatrix}
\ee
Assuming that the triplets $\psi$ and $\psi'$ are Standard Model fermions
(for Yukawa terms $\psi' = \psi^c$ and for Majorana terms
$\psi' = \psi$)
while the sextet $\tilde\chi$ is a flavon field, the Yukawa and Majorana
couplings $Y^{Yuk},Y^{Maj}$ are generated when the components of $\tilde\chi$
develop their VEVs. More intuitively, one can regard the matrix
$\hat\chi$ as a two-index symmetric tensor field in the sextet representation
which obtains a VEV $\langle \hat\chi \rangle$.

For practical purposes, especially when constructing and evaluating
the sextet flavon potential, it is however technically easier to work with the
sextet being a six component column vector. Furthermore, as the real basis is
more convenient, we define the unitary transformation
\be
\chi ~=~ \tilde{\m R} \, \tilde\chi \ ,\label{trafo2real}
\ee
with
\be
\tilde{\m R} ~=~ \frac{(1+i)}{6 \sqrt{2}}
\begin{pmatrix}
-4 & 2 & 2 & -\sqrt{2} & 2 \sqrt{2} & -\sqrt{2} \\
0 & 2i\sqrt{3}&-2i\sqrt{3}&i\sqrt{6}&0&-i\sqrt{6} \\
0&-i\sqrt{3}\,b_7&i\sqrt{3}\,b_7&i\sqrt{6}\,b_7&0&-i\sqrt{6}\,b_7 \\
0&-\sqrt{3}\,b_7&-\sqrt{3}\,b_7&-\sqrt{6}\,b_7&0&-\sqrt{6}\,b_7 \\
 -2 \sqrt{2} & -2 \sqrt{2} & -2 \sqrt{2} &2&2&2\\
i\sqrt{6}\,\bar b_7&0&0&0&2i\sqrt{3}\,\bar b_7 &0
\end{pmatrix} .
\ee
Then the generators of the sextet take the form
\bea
\m S&=&\tilde{\m R} \, \tilde{\m S} \, \tilde{\m R}^{\dagger}  ~=~
\text{Diag}\, (-1\,,\,-1\,,\,1\,,\,1\,,\,1\,,\,1) \  , \\[4mm]
\m T&=&\tilde{\m R} \, \tilde{\m T} \, \tilde{\m R}^{\dagger} ~=~ \frac{1}{2}
\begin{pmatrix}
1&1&0&0&\sqrt{2}&0 \\
-1&-1&0&0&\sqrt{2}&0 \\
0&0&-1&-\sqrt{3}&0&0 \\
0&0&\sqrt{3}&-1&0&0 \\
\sqrt{2}&-\sqrt{2}&0&0&0&0 \\
0&0&0&0&0&2 \\
\end{pmatrix} , \\[4mm]
\m U&=&\tilde{\m R} \, \tilde{\m U} \, \tilde{\m R}^{\dagger}  ~=~
\text{Diag}\, (1\,,\,-1\,,\,-1\,,\,1\,,\,1\,,\,1) \ ,\\[4mm]
\m V&=&\tilde{\m R} \, \tilde{\m V} \, \tilde{\m R}^{\dagger} ~=~ \frac{1}{6}
\begin{pmatrix}
6&0&0&0&0&0 \\
0&0&6&0&0&0 \\
0&6&0&0&0&0 \\
0&0&0&4&\sqrt{6}&-\sqrt{14} \\
0&0&0&\sqrt{6}&3&\sqrt{21} \\
0&0&0&-\sqrt{14}&\sqrt{21}&-1 \\
\end{pmatrix} ,
\eea
while the matrix $\tilde{\m C}$ changes to
\be
\tilde{\m R}^\ast \, \tilde{\m C} \, \tilde{\m R}^{\dagger} ~=~
\mathds{1}^{}_{6\times 6}  \ .
\ee




\section{\label{towards}Towards a realistic model}
\cleqn

As we are working in the real basis for the sextet from now on, we should
express $\tilde \chi$ in the matrix of Eq.~(\ref{yukmatrix}) in
terms of $\chi$, see Eq.~(\ref{trafo2real}). The result is\footnote{Note that
  we are not changing the basis for the triplets. Therefore the
  matrix $\hat\chi$ of Eq.~(\ref{yukmatrix}) is identical to the one of Eq.~(\ref{yukmatrixreal}).}
\bea
\hat\chi&\!\!=\!\!& - \frac{(1+i)}{6\sqrt{2}}
\left[
\chi_1 \begin{pmatrix}
4&1&1\\1&-2&-2\\1&-2&-2
\end{pmatrix}
-i\sqrt{3}\,\chi_2 \begin{pmatrix}
0&1&-1\\1&2&0\\-1&0&-2
\end{pmatrix}
-i\sqrt{3}\,b_7 \,\chi_3 \begin{pmatrix}
0&1&-1\\1&-1&0\\-1&0&1
\end{pmatrix}
\right.\nonumber\\[2mm]
&&+ \left.\sqrt{3}\,b_7 \,\chi_4 \begin{pmatrix}
0&1&1\\1&1&0\\1&0&1
\end{pmatrix}
+ \sqrt{2} \,\chi_5 \begin{pmatrix}
2&-1&-1\\-1&2&-1\\-1&-1&2
\end{pmatrix}
- i \sqrt{6}\,\bar b_7 \,\chi_6 \begin{pmatrix}
1&0&0\\0&0&1\\0&1&0
\end{pmatrix}
\right] .\label{yukmatrixreal}
\eea
In this form it is evident how the vacuum structure of a sextet flavon field
can give rise to the Yukawa and Majorana matrices $Y^{Yuk},Y^{Maj}$.
Using the sextet flavons, the (third family) effective Yukawa couplings are
supposed to originate from terms of the form, 
\be
{\mathcal L}^{Yuk} \sim \psi^{}_{i}\hat{\chi}_{ij}^{Yuk}\psi_{j}^{c}H \ ,
\label{opYuk-2}
\ee
while the effective Majorana couplings will originate from terms like
\be
{\mathcal L}^{Maj} \sim \psi^{}_{i}\hat{\chi}_{ij}^{Maj}\psi^{}_{j}HH \ ,
\label{opMaj-2}
\ee
where both terms are invariant under $PSL_2(7)$, with the indices $i,j$ now
labelling the entries of the matrix in Eq.~(\ref{yukmatrixreal}).
In Eqs.~(\ref{opYuk-2}) and (\ref{opMaj-2}) we have introduced two different sextet flavons,
namely $\hat{\chi}_{ij}^{Yuk}$ and $\hat{\chi}_{ij}^{Maj}$, where the VEV of
$\hat{\chi}_{ij}^{Yuk}$ must be aligned along the third family direction,
$\langle \hat{\chi}^{Yuk}_{ij} \rangle = V\delta_{i3}\delta_{j3}$,
while $\hat{\chi}_{ij}^{Maj}$ must develop VEVs which will give rise to the 
TB neutrino mass matrix, $\langle \hat{\chi}^{Maj}_{ij} \rangle \propto ({M^{\nu}_{TB}})_{ij}$.



\subsection{Sextet Vacuum Alignment}


\subsubsection{The alignment of ${\bs{\hat{\chi}_{ij}^{Maj}}}$}
In order to achieve the desired alignment for the VEV of $\hat{\chi}_{ij}^{Maj}$,
we require its components $\chi^{Maj}_{a}$, where $a = 1, \ldots , 6$,
to have an alignment of the form
\be
\langle \chi^{Maj}_{TB} \rangle ~=~ - \, \frac{ 2\,\sqrt{6}}{(1+i)} \cdot (0\,,\,0\,,\,0\,,\,  \alpha_4 \,,\,
\alpha_5 \,,\,\alpha_6) \ , \label{TBalignment}
\ee
which is left unchanged by the action of the $PLS_2(7)$ sextet generators $\m
S$ and $\m U$. This is clear from Eq.~(\ref{yukmatrixreal}) where it is observed that
the matrices proportional to $\chi_{4,5,6}$ all satisfy the TB conditions in Eq.~(\ref{abc}).
Consequently, the resulting Majorana matrix $Y^{Maj}_{TB}$ is
invariant under $S$ and $U$:
$$
S\cdot Y^{Maj}_{TB}\cdot  S^T ~=~Y^{Maj}_{TB} \ , \qquad
U\cdot Y^{Maj}_{TB}\cdot  U^T ~=~Y^{Maj}_{TB} \ .
$$
Since $S$ and $U$ are the symmetries of the tri-bimaximal mass matrix,
$Y^{Maj}_{TB}$ is diagonalised by the tri-bimaximal mixing matrix
$U_{TB}$. We find the following neutrino mass eigenvalues,
\bea
m_1&=&- b_7 \, \alpha_4 +  \sqrt{6} \,\alpha_5 -i \sqrt{2}\,\bar b_7
\,\alpha_6 \ , \\
m_2&=& 2 b_7 \, \alpha_4 -i \sqrt{2}\,\bar b_7
\,\alpha_6 \ , \\
m_3&=& b_7 \, \alpha_4 + \sqrt{6} \,\alpha_5 +i \sqrt{2}\,\bar b_7
\,\alpha_6 \ .
\eea

It is interesting to note that an alignment in which only the sixth component
of the sextet has a non-vanishing value, leads to a Majorana matrix that is
additionally symmetric under $T$. Actually, any $PSL_2(7)$ sextet of the form
$$
\chi^{[{\bf 1}]}~=~(0\,,\,0\,,\,0\,,\,0\,,\,0\,,\,\chi_6) \ ,
$$
is left invariant under the subgroup $S_4$. Giving a VEV
to this field breaks $PSL_2(7)$ down to $S_4$. The complete sextet
decomposes as
$$
{\bf 6} ~\rightarrow ~ {\bf 1} + {\bf 2} + {\bf 3'} \ ,
$$
where ${\bf 3'}$ is the triplet of $S_4$ for which $\det (U) = -1$.
Since in our real basis the matrices $\m S$, $\m T$, $\m U$ are ``quasi''
block diagonal, it is easy to check that the directions of the sextet that
become the doublet and the triplet are
\bea
\chi^{[{\bf 2}]} ~=~ (0\,,\,0\,,\,\chi_3\,,\,\chi_4\,,\,0\,,\,0) \ , \qquad
\chi^{[{\bf 3'}]} ~=~ (\chi_1\,,\,\chi_2\,,\,0\,,\,0\,,\,\chi_5\,,\,0) \ .
\notag
\eea
If we want the VEVs of these fields to be unaltered by $\m S$ and $\m U$, we
must set $\langle \chi_1\rangle =\langle\chi_2\rangle=\langle\chi_3\rangle=0$
which brings us back to the structure of $\langle \chi^{Maj}_{TB}\rangle$ of
Eq.~(\ref{TBalignment}).


\subsubsection{The alignment of ${\bs{\hat{\chi}_{ij}^{Yuk}}}$}
In order to generate a Yukawa matrix which gives mass to only the third
generation (top quark, bottom quark or tau lepton),
we require its components $\chi^{Yuk}_{a}$, where $a = 1, \ldots , 6$,
to have an alignment of the form
\be
\langle \chi^{Yuk}_{\mathrm{top}} \rangle ~\propto ~
(1-i) \cdot  (-1~,~-i \sqrt{3}~,~ i\sqrt{3}/b_7 ~,~ \sqrt{3}/b_7~,~\sqrt{2} ~,~0) \ .\label{topalignment}
\ee
With this vacuum structure for the sextet $\chi^{Yuk}_{\mathrm{top}}$, one can
easily show that only the $(3,3)$-entry of the Yukawa matrix in
Eq.~(\ref{yukmatrixreal}) gets filled in.


\subsubsection{The sextet vacuum alignment problem}
To obtain alignments of the type $\langle \chi^{Maj}_{TB} \rangle $ of
Eq.~(\ref{TBalignment}) and $\langle \chi^{Yuk}_{\mathrm{top}}\rangle  $ of
Eq.~(\ref{topalignment}) it is necessary to study $PSL_2(7)$ symmetric
potentials for the flavon sextet. Even though a detailed analysis of the
dynamics within our framework goes well beyond the scope of the present
article, we want to illustrate our procedure which proves successful in
finding suitable sextet potentials~\cite{future}. Following along the lines of
\cite{delta27-LQ:King,Z7Z3-LQ} we consider sextet potentials of
the form
\be
V~=~-m^2 \cdot \sum_{i=1}^6 \, \chi^\dagger_i \chi^{}_i ~+ \,
\sum_{\alpha=0}^5 \kappa^{}_\alpha \cdot \!\!\!
\sum_{i,j,k,l=1}^6   c_{ijkl}^{\alpha} \,  \chi^\dagger_i
\chi^{}_j \chi^\dagger_k \chi^{}_l  \ .
\ee
In general, the quartic part consists of several independent invariants which are
denoted by~$\alpha$. The coefficients $c_{ijkl}^\alpha$ are defined by the
Clebsch Gordan coefficients of the $PSL_2(7)$ Kronecker products. As we have
two identical sextet fields together with their complex conjugates, it is
sufficient to consider the square of the symmetric quadratic product
$$
{\bf (6\otimes 6)}_s \otimes {\bf (6\otimes 6)}_s ~=~ {\bf (1 ~+~6~+~6~+~8)}
\otimes {\bf (1 ~+~6~+~6~+~8)} \ ,
$$
which yields a total of six independent invariants, two of which are also
$SU(3)$ invariant.\footnote{A similar potential built out of $PSL_2(7)$
  (anti-)triplet flavons yields quartic invariants which are identical to
  those of $SU(3)$ because the product ${\bf 3 \otimes   \bar 3 = 1 + 8}$ has
  identical Clebsch Gordan coefficients in both cases. With solely one flavon
  field, the only quartic invariant would be the square of the quadratic one,
  $\bar{\phi^i}^\dagger \bar \phi_i \bar{\phi^j}^\dagger \bar \phi_j$, so that
  it is impossible to obtain an alignment for a single (anti-)triplet flavon.}
Their explicit derivation
is straightforward, though tedious since it requires the knowledge of all
Clebsch Gordan coefficients for the product $({\bf 6 \otimes 6})_s$. Among
others, this necessitates the construction of the $PSL_2(7)$ octet in the
$\m S^{[{\bf 8}]},\m T^{[{\bf 8}]},\m U^{[{\bf 8}]},\m V^{[{\bf 8}]}$ basis.

Having calculated the coefficients $c_{ijkl}^\alpha$, the
discrete family symmetry does not fix the weights $\kappa_\alpha$ with which
each invariant enters in the potential. Therefore, from the model building
point of view, we can choose them so that the desired alignment vectors
minimise~$V$. To this end, we need to evaluate the first derivatives of each
of the six invariants for a particular vacuum alignment. Requiring vanishing
first derivatives already constrains the possibilities for
$\kappa_\alpha$. Next the matrix of second derivatives, i.e. the Hessian, must
be calculated. Only in the case where the Hessian is positive  definite, the
potential is minimised by the desired alignment. In both cases, i.e. for
$\langle \chi^{Maj}_{TB} \rangle$ and $\langle \chi^{Yuk}_{\mathrm{top}}
\rangle$, we can
find parameter ranges for $\kappa_\alpha$ which lead to a suitable potential.
In hindsight one could then argue that a particular messenger sector is
responsible for the suppression or absence of some of the six independent
invariants.



\subsection{Other Yukawa operators}

The complete model should include, in addition to the
third family Yukawa couplings, also the other operators responsible for the
Yukawa couplings of the first and second family of quarks and leptons.
In order to do this, one may construct the remaining parts of the model
along the lines of the model in \cite{delta27-LQ:King}
by extending the Yukawa sector to include anti-triplet flavons $\bar{\phi}$,
with the leading order operators given by,
\be
{\mathcal L}^{Yuk} ~\sim~ \psi^{}_{i}\hat{\chi}^{Yuk}_{ij}\psi_{j}^{c}H
~+~ \bar{\phi}^{i}\psi^{}_{i}\bar{\phi}^{j}\psi_{j}^{c}H ~+\,\ldots ~\ ,
\label{Yukawa}
\ee
where the term involving the sextet $\hat{\chi}^{Yuk}_{ij}$, with the vacuum
alignment discussed above, leads to the third family Yukawa
couplings, while terms involving anti-triplet flavons generate the first and
second family Yukawa couplings. These terms may originate from diagrams such
as those in Figs.~\ref{diagrams}($a,b$), leading to a realistic quark and
lepton mass spectrum.




\section{\label{conclusion}Conclusion}
\cleqn

We have argued that, in the quest for a candidate discrete family
symmetry $G_f$ suitable for combining with $SO(10)$ type
unification, one should seek groups with the following features:
($i$) $G_f$ should have complex triplet and (real or complex)
sextet representations, and ($ii$) $S_4 \subset G_f$. We have
pointed out that the projective special linear group $PSL_2(7)$, also known as
$\Sigma(168)$,  
satisfies both requirements, and furthermore is the unique simple
finite group containing complex triplets. This
makes $PSL_2(7)$ ideally suitable as a discrete family symmetry
for describing quark and lepton masses and mixing in the framework
of $SO(10)$ type unified models.

We have shown how the flavon fields in the sextet representation
of ${PSL_2(7)}$ play a crucial role both in obtaining
tri-bimaximal neutrino mixing as well as in generating the third
family charged fermion Yukawa couplings. In preparation for
physical applications, we have derived the triplet representation
of ${PSL_2(7)}$ in the basis $S,T,U,V$ where $S,T,U$ are the
familiar triplet generators of $S_4$ in the diagonal charged
lepton basis where $T$ is diagonal. We have also derived a
convenient real sextet representation, responsible for the
Majorana couplings of the neutrino operators and the Yukawa
couplings of the third family charged fermions. We have identified
the sextet vacuum alignment which preserves the sextet generators
$\m S,\m U$ and lead to tri-bimaximal neutrino mixing. We have
also displayed the sextet vacuum alignment which breaks all the
generators and gives rise to the large third family charged
fermion Yukawa couplings. We have briefly described our approach
to the vacuum alignment problem, which will give rise to both
these alignments.

In conclusion, we have proposed a new and promising class of
 $G_f~\otimes$~GUT models based on $PSL_2(7)$ family
symmetry and $SO(10)$ type unification, focusing on the symmetry
aspects of such a model and in particular on the new features
arising from the presence of flavons in the sextet representation.
A complete dynamical model of this kind based on ${{PSL_2(7)}}=\Sigma(168)$,
including a full analysis of the vacuum alignments of sextet and
anti-triplet flavons, is in progress and will be presented
elsewhere~\cite{future}.

~\\[5mm]
\noindent{\bf{\Large{Acknowledgments}}}\\

\noindent
SFK acknowledges support from 
STFC Rolling Grant ST/G000557/1 and
EU Network MRTN-CT-2004-503369.
The work of CL is supported by the STFC Rolling Grant ST/G000557/1.




\end{document}